\documentclass[10pt,twocolumn,letterpaper]{article}

\usepackage{cvpr}              

\usepackage{utfsym}
\usepackage{multicol}
\usepackage{multirow}
\usepackage{threeparttable}
\usepackage{pifont}
\graphicspath{{./Imgs/}}
\DeclareGraphicsExtensions{.pdf,.png,.jpg}

\newcommand{\figref}[1]{Fig.~\ref{#1}}
\newcommand{\tabref}[1]{Table~\ref{#1}}

\usepackage[accsupp]{axessibility} 

\def\ie{\textit{i.e.,~}}

\def\etal{\textit{et al.}~}
\def\sArt{{state-of-the-art~}}

\definecolor{cvprblue}{rgb}{0.21,0.49,0.74}
\usepackage[pagebackref,breaklinks,colorlinks,citecolor=cvprblue]{hyperref}

\def\paperID{5502} 
\def\confName{CVPR}
\def\confYear{2025}

\title{Cross-Modal Interactive Perception Network with Mamba for Lung Tumor Segmentation in PET-CT Images}

\author{
    Jie Mei\textsuperscript{1,2}\footnotemark[1],
    Chenyu Lin\textsuperscript{2}\footnotemark[1],
    Yu Qiu\textsuperscript{3}\footnotemark[2],
    Yaonan Wang\textsuperscript{1},
    Hui Zhang\textsuperscript{1},
    Ziyang Wang\textsuperscript{4},
    Dong Dai\textsuperscript{4}
     \\
    \small
    \textsuperscript{1}Hunan University \quad
    \textsuperscript{2}Nankai University \quad
    \textsuperscript{3}Hunan Normal University \quad
    \textsuperscript{4}Tianjin Medical University Cancer Institute and Hospital \\
    \small
    \texttt{\{jiemei, yaonan, zhanghui1983\}@hnu.edu.cn}, 
    \texttt{\{chenyulin, yqiu\}@mail.nankai.edu.cn}
    \\
    \small
    \texttt{\{wangziyang, dongdai\}@tmu.edu.cn}
}

\begin{document}
\maketitle

\footnotetext[0]{$^*$Equal contribution. \quad $^{\dagger}$Corresponding author.}

\begin{abstract}
Lung cancer is a leading cause of cancer-related deaths globally. PET-CT is crucial for imaging lung tumors, providing essential metabolic and anatomical information, while it faces challenges such as poor image quality, motion artifacts, and complex tumor morphology. Deep learning-based models are expected to address these problems, however, existing small-scale and private datasets limit significant performance improvements for these methods. 
Hence, we introduce a large-scale PET-CT lung tumor segmentation dataset, termed PCLT20K, which comprises 21,930 pairs of PET-CT images from 605 patients. 
Furthermore, we propose a cross-modal interactive perception network with Mamba (CIPA) for lung tumor segmentation in PET-CT images. Specifically, we design a channel-wise rectification module (CRM) that implements a channel state space block across multi-modal features to learn correlated representations and helps filter out modality-specific noise. A dynamic cross-modality interaction module (DCIM) is designed to effectively integrate position and context information, which employs PET images to learn regional position information and serves as a bridge to assist in modeling the relationships between local features of CT images. Extensive experiments on a comprehensive benchmark demonstrate the effectiveness of our CIPA compared to the current state-of-the-art segmentation methods. We hope our research can provide more exploration opportunities for medical image segmentation.
The dataset and code are available at \url{https://github.com/mj129/CIPA}.
\end{abstract}
    
\section{Introduction}

Lung cancer remains one of the most significant causes of cancer-related mortality globally. Positron Emission Tomography-Computed Tomography (PET-CT) has become a cornerstone in the imaging and evaluation of lung tumors, providing crucial metabolic and anatomical information \cite{xiang2022modality}. The accurate segmentation of lung tumors in PET-CT images is essential for effective diagnosis, treatment planning, and monitoring of therapeutic responses. 
However, this task faces substantial challenges due to the complex nature of lung tumors and the variability in PET-CT images, such as poor image quality, motion artifacts or noise, complex morphology, irregular shape, and unclear tumor boundaries. Deep learning-based segmentation models are expected to address these problems, which can significantly improve segmentation accuracy, reduce the workload of radiologists, and lead to consistent assessments.

\begin{figure}[!tb]
\centering
\includegraphics[width=0.86\linewidth]{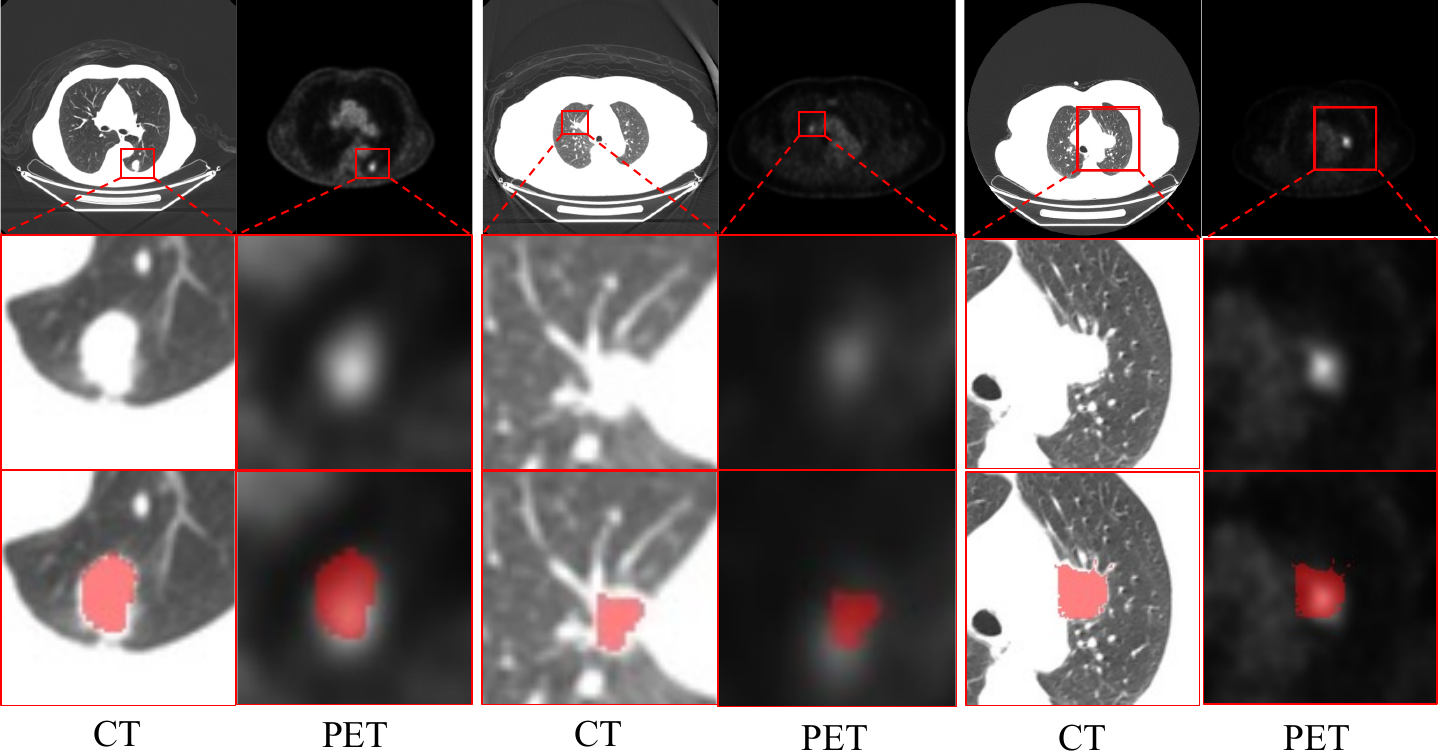}
\caption{Some examples of the PET-CT images and the corresponding tumor picked from the PCLT20K dataset. The metabolic data from PET enhances sensitivity to lesion location, while the anatomical details from CT help precise localization and morphological characterization.
}
\label{fig:example}
\end{figure}

However, these models are usually data-hungry, and their performance gains mainly benefit from the large amounts of accurately labeled training data. Although some annotated datasets have been build for model training and validation, they are usually small-scale \cite{zhong20183d,zhong2019simultaneous,kumar2020co,li2020deep,fu2021multimodal,xiang2022modality,wang2023automated,chen2023pet}. The two largest publicly reported datasets of lung only comprise 6,985 PET-CT pairs from 126 patients \cite{xiang2022modality} and 3,905 pairs from 290 patients \cite{wang2023automated}, respectively. What's worse, almost all of these datasets are private, severely hindering further advancements in this field. We summarize the recent PET-CT tumor segmentation datasets in \tabref{tab:datasets}.
Although these datasets have supported the development of some segmentation models \cite{fu2021multimodal}, their limited scale and restricted accessibility significantly constrain the models' applicability and robustness in broader clinical scenarios. Consequently, the foremost challenge for PET-CT lung tumor segmentation lies in the absence of a publicly available, large-scale dataset. This limitation arises from stringent privacy concerns associated with PET-CT data and the need for substantial clinical expertise to ensure accurate annotation.

To address these limitations, we first establish a comprehensive, large-scale PET-CT lung tumor segmentation dataset, named PCLT20K. PCLT20K consists of 21,930 pairs of PET-CT images collected from 605 unique patients.
All PET-CT image pairs are manually annotated with the pixel-level masks of tumors. To ensure the quality and reliability of these annotations, a rigorous three-stage annotation process is conducted with all annotators being experienced clinicians. Representative examples of PET-CT images and their corresponding tumor labels are shown in \figref{fig:example}, illustrating the significant challenges of PET-CT lung tumor segmentation: 1) the low contrast, low spatial resolution, and substantial noise can obscure the tumor appearances, especially their boundaries; 2) the shapes and structures of tumors are irregular and complex; 3) tumors vary considerably in size, location, and appearance, with some tumors being very small or situated in intricate anatomical regions; and 4) some lung tumors grow infiltratively, diffusely invading surrounding tissues. These factors are still challenging for most existing segmentation algorithms.

To achieve high-performance lung tumor segmentation from PET-CT images, we propose a cross-modal interactive perception network with Mamba (named CIPA) for lung tumor segmentation in PET-CT images. Specifically, a channel-wise rectification module (CRM) is proposed by implementing a channel state space block across multi-modal features, which enhances shared representation learning and helps to filter out modality-specific noise by emphasizing the correlated features between the two modalities.
Additionally, considering that PET images often reveal potential lesion areas while CT images present detailed texture information of the lungs, we design a dynamic cross-modality interaction module (DCIM)
to effectively integrate position and context information. DCIM employs the interactive region Mamba block and local Mamba block to enable effective cross-modality fusion by dynamically interacting PET and CT information, allowing for a more comprehensive and synergistic feature representation.

To summarize, our main contributions are as follows:
\begin{itemize}
    \item We propose the first publicly available large-scale dataset for researching the task of lung tumor segmentation from PET-CT images, named \textbf{PCLT20K}, which contains 21,930 pairs of PET-CT images from 605 patients with high-quality pixel-level annotations.
    \item We propose a cross-modal interactive perception network with Mamba, named \textbf{CIPA}, which designs a channel-wise rectification module and a dynamic cross-modality interaction module to effectively interact with the correlation information between multi-modal data.
    \item Based on the PCLT20K dataset, we establish a comprehensive benchmark to facilitate future research. Extensive experiments validate that our CIPA can achieve state-of-the-art performance.
\end{itemize}

\begin{table}[!tb]
    \centering
    \footnotesize
    \renewcommand{\arraystretch}{1}
        \begin{tabular}{c|c|c|c|c}
        \hline
            Dataset & Location & \#Cases & \#Pairs & Public
            \\ \hline
            HECKTOR
            \cite{oreiller2022head} & Head \& Neck & 845 & - & \ding{52}
            \\
            MFCNet \cite{wang2023mfcnet} & Pancreas & 93 & - & \ding{55}
            \\
            AutoPET \cite{gatidis2023autopet} & Whole Body & 900 & - & \ding{52}
            \\\hline
            STS \cite{vallieres2015radiomics} & Lung & 51 & 2,409 & \ding{52}
            \\
             Co-Segmentation \cite{zhong20183d} & Lung & 32 & - & \ding{55}
            \\
            DFCN \cite{zhong2019simultaneous} & Lung & 60 & - & \ding{55}
            \\
            Co-Learning \cite{kumar2020co} & Lung & 50 & 855 & \ding{55}
            \\
            FVM-PET \cite{li2020deep} & Lung & 36 & - & \ding{55}
            \\
            NSCLC \cite{fu2021multimodal} & Lung & 50 & 876 & \ding{55}
            \\
            MoSNet \cite{xiang2022modality} & Lung & 126 & 6,985 & \ding{55}
            \\
            Dual-Modality \cite{wang2023automated} & Lung & 290 & 3,905 & \ding{55}
            \\
            Joint-Level-Set \cite{chen2023pet} & Lung & 20 & - & \ding{55}
            \\\hline
            PCLT20K (Ours) & Lung & 605 & 21,930 & \ding{52}
            \\ \hline
        \end{tabular}
    \caption{Summary of PET-CT tumor segmentation datasets.}
    \label{tab:datasets}
\end{table}

\section{Related Work}

\subsection{Tumor Segmentation in PET-CT Images}

The availability of high-quality datasets is crucial for developing and evaluating segmentation algorithms \cite{qiu2022miniseg}. In recent years, some PET-CT tumor segmentation datasets have been proposed successively, covering organs such as the lung \cite{angelus2020large,kumar2020co}, pancreas \cite{wang2023mfcnet}, head and neck \cite{oreiller2022head}, and so on \cite{gatidis2023autopet}. However, as summarized in \tabref{tab:datasets}, almost all existing lung tumor segmentation datasets are small-scale and private \cite{fu2021multimodal,xiang2022modality,wang2023automated,chen2023pet}, which limits the further development of the field. Despite these limitations, several lesion segmentation algorithms based on these datasets have been proposed \cite{qiu2022delving}. Convolutional neural networks (CNNs), especially fully convolutional networks (FCNs), have been widely applied to tumor segmentation from PET-CT images and remarkably boosted the segmentation performance \cite{kumar2020co,mei2021sanet}. 
Among these methods, UNet and its variants demonstrated superior accuracy on PET-CT datasets thanks to their enhanced feature extraction capabilities \cite{xiang2022modality}. More recently, some works have employed vision Transformers for PET-CT image segmentation \cite{zhao2024mmca}. These models leverage self-attention mechanisms to capture global contextual information, outperforming traditional FCN-based models, particularly in cases with complex tumor morphologies. Besides, there are efforts dedicated to using multi-modal fusion techniques to further improve PET-CT segmentation accuracy \cite{wang2023mfcnet}.
Nevertheless, the small scale and privacy of current PET-CT datasets limit the reproducibility of these algorithms and impede the advancement of this field.

\subsection{Mamba in Medical Image Segmentation}

The emergence of Mamba \cite{gu2023mamba} has brought new possibilities to the field of deep learning-based medical image segmentation. Traditional CNN-based models are inherently limited in their ability to capture long-range dependencies. Transformers ensure long-range modeling through self-attention mechanisms, while they do so at the cost of quadratic time complexity. Mamba addresses this issue by introducing a Selective State Space Model (SSM) that achieves long-range interaction with linear time complexity, which achieves the current optimal performance \cite{gu2023mamba}. U-shaped Mamba networks have become the mainstream architectures, such as U-Mamba \cite{ma2024u}, SegMamba \cite{xing2024segmamba}, and Swin-UMamba \cite{liu2024swin}.
Huang \etal \cite{huang2024localmamba} introduced a novel local scanning strategy that divides images into distinct windows, effectively capturing local dependencies while maintaining a global perspective. Mamba-based models have also been extended to lightweight \cite{liao2024lightm}, semi-supervised \cite{ma2024semi}, and weakly-supervised \cite{wang2024weak} medical image segmentation field.
Despite their advancements, Mamba struggles to effectively explore regional correlations and complementary information between multi-modal images due to its 2D selective scanning mechanism, thereby restricting the performance of tumor segmentation in PET-CT images.

\section{PCLT20K Dataset}
\label{sec:dataset}
\subsection{Data Collection}
\label{ssec:DataCollection}

The PET-CT images of PCLT20K are collected from the department of molecular imaging and medicine at a top-tier hospital. The patients' examination dates range from June 2016 to April 2020. Besides, the PET-CT images are acquired using a GE Discovery Elite PET/CT scanner (GE Medical Systems). Prior to PET/CT imaging, patients fast for approximately six hours with a serum glucose level maintained below 11.1 mmol/L. Images are acquired 50 to 60 minutes after the injection of 4.2 MBq/kg $^{18}$F-FDG. A spiral CT scan (80 mAs, 120 kVp, and 5-mm slice thickness) is performed for precise anatomical localization and attenuation correction, and a PET emission scan in 3D mode is followed from the distal femur to the top of the skull. 
The voxel size of CT images is $0.98 \times 0.98 \times 2.8$ mm.
PET images are reconstructed using an ordered-subset expectation maximization (OSEM) iterative algorithm to achieve a final pixel size of $3.6 \times 3.6 \times 3.3$ mm.
It is worth noting that during data collection at the hospital, all private information is removed, including hospital names, doctor names, patient names, and other identifying details. Additionally, images with foreign object interference and motion artifacts are excluded to ensure the quality of the collected images.

\subsection{Data Annotation}
\label{ssec:DataAnnotation}

To guarantee the accuracy and reliability of tumor region annotations, we employ a three-stage annotation process conducted by experienced physicians. The first annotation stage occurs when doctors diagnose the patients in the hospital. During this stage, doctors analyze CT and PET images to generate examination reports that include the approximate location, appearance, and type of the tumor. Referring to these medical reports, physicians in the second stage conduct a detailed annotation for each case on a slice-by-slice basis, marking all the pixels in the tumor regions. In the third annotation stage, another doctor reviews and refines the annotations to form the final pixel-level annotations of the tumor regions. If there are any discrepancies between the annotations from the second and third stages, the doctors will discuss and re-annotate the tumor regions together.
Note that in the second and third stages, doctors select the tumor parenchyma area as the lesion area for annotation. Due to the uneven growth of the tumor and its invasion into surrounding tissues, the annotated lesion area may be discontinuous.

We finally obtain 21,930 2D slice pairs of PET-CT images from 605 unique patients with high-quality pixel-level annotations. We randomly split them into a training set and a testing set at an 8:2 ratio, naming them PCLT20K-TR (17,416 pairs) and PCLT20K-TE (4514 pairs), respectively. The data splitting is conducted at the patient level.

In the PCLT20K dataset, PET-CT images undergo preprocessing before being input into the network.
Based on doctors' experience, the Hounsfield Unit (HU) values of the CT images are clipped to the range [-1200, -200] and then normalized to $[0, 255]$.
The PET images are normalized by converting them to Standard Uptake Value (SUV) \cite{thie2004understanding}, which are then scaled to $[0, 255]$.
All PET-CT slices in PCLT20K are resized to the same resolution of $512\times 512$ to ensure they share the same coordinate space, which is a standard procedure for analyzing PET-CT data \cite{xiang2022modality, kumar2019co}.

\begin{figure}[!tb]
\centering
\subfloat[]{\includegraphics[width=.46\linewidth] {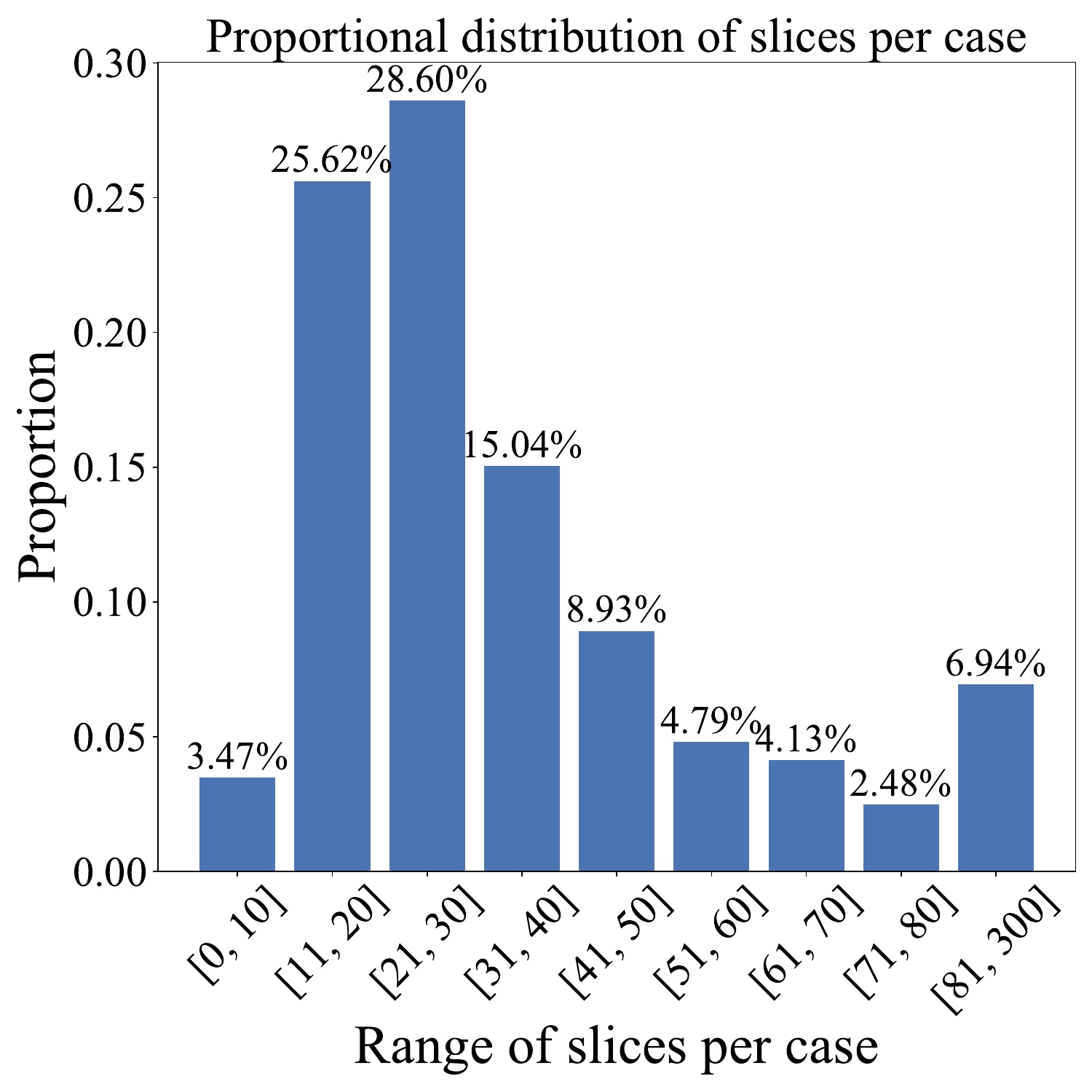}\label{fig:statistic_num_slice}} \hspace{0.1in}
\subfloat[]{\includegraphics[width=.41\linewidth]{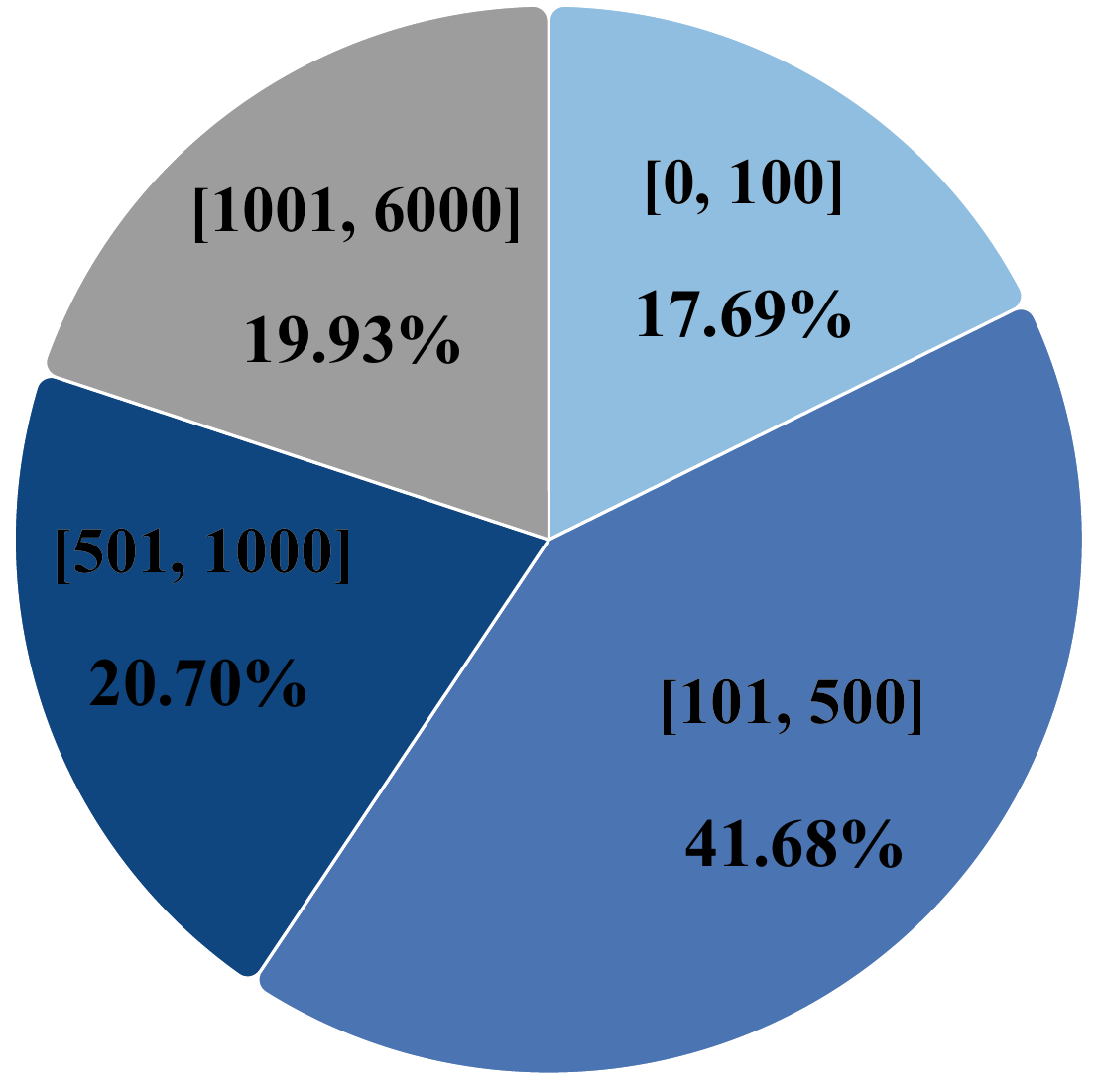}\label{fig:statistic_size_chart}}
\caption{(a) Proportional statistics of the number of slices per case. (b) Statistics of the pixel counts in tumor regions.}
\label{fig:statistic}
\end{figure}

\begin{figure}[!tb]
\centering
\subfloat[]{\includegraphics[width=.43\linewidth] {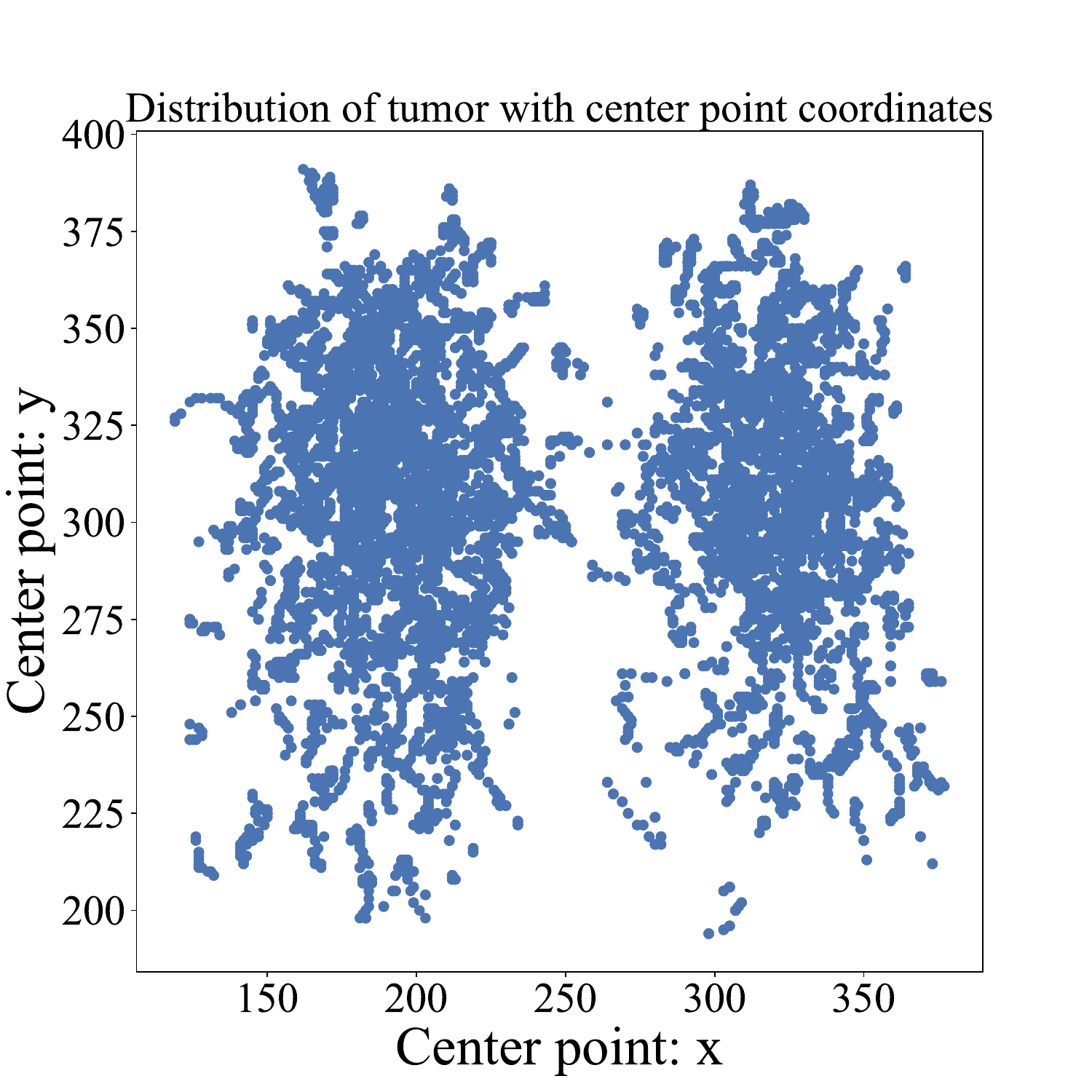}\label{fig:statistic_center_mask}} \hspace{0.1in}
\subfloat[]{\includegraphics[width=.43\linewidth]{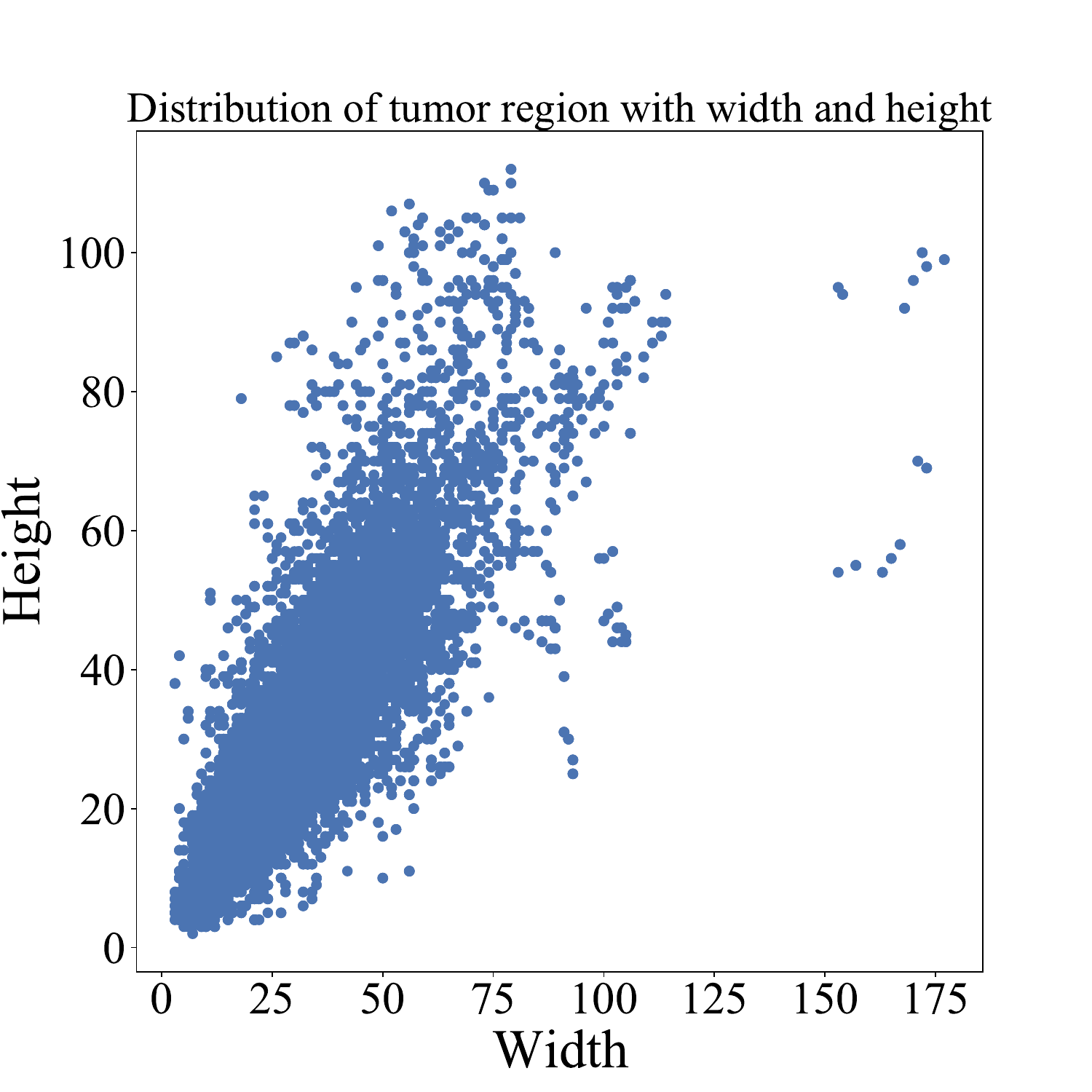}\label{fig:statistic_size_mask}}
\caption{Distribution of tumor center points and sizes. (a) Distribution of tumor in terms of center point coordinates. (b) Distribution of tumor in terms of width and height.}
\label{fig:attributes}
\end{figure}

 \begin{figure*}[ht]
\centering
\includegraphics[width=0.98\linewidth]{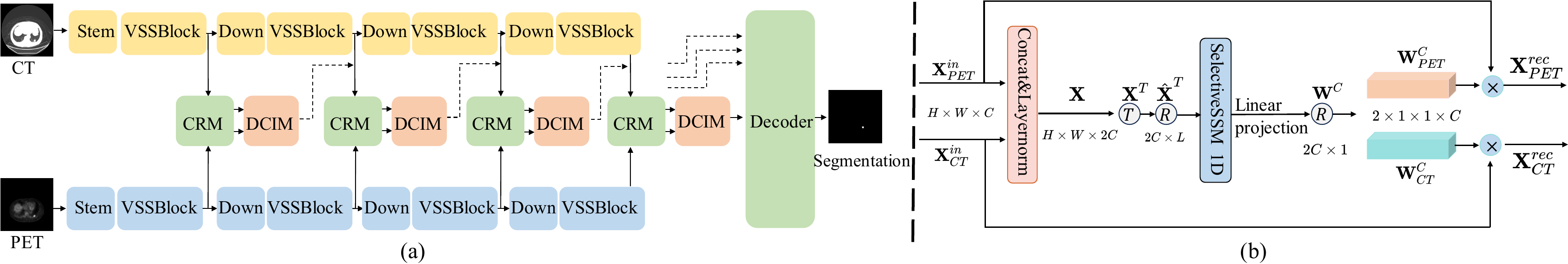}
\caption{(a) Overall architecture of our proposed cross-modal interactive perception network with Mamba (CIPA) for lung tumor segmentation in PET-CT images. CIPA consists of: (1) a channel-wise rectification module (CRM) to learn shared representations; (2) a dynamic cross-modality interaction module (DCIM) to integrate position and context information effectively. (b) Illustration of CRM.
}
\label{fig1}
\end{figure*}

\subsection{Data Statistics}
\label{ssec:DataStatistics}
We conduct statistical analyses to present the distribution of tumor sizes and locations in the PCLT20K dataset.
The proportional statistics of the number of slices per case are shown in \figref{fig:statistic}(a), 72.73\% of the tumors have fewer than 40 slices, while only 13.55\% of the tumors exhibit more than 60 slices. Notably, PCLT20K demonstrates considerable variability, with the minimum number of tumor-containing slices being 4 and the maximum reaching 278.
The size distribution of tumors, as visualized in \figref{fig:statistic}(b), reveals that a majority of tumors occupy relatively limited spatial regions within the slices.
In detail, 59.37\% of tumors have an area smaller than 500 pixels. The number of tumors whose sizes are over 1,000 pixels only accounts for 19.93\%, reflecting the distribution's skew toward smaller lesions within the PCLT20K dataset. Among all lesions, the smallest tumor only has 11 pixels, while the largest occupies an area of 5,830 pixels. Furthermore, we plot the locations of center points of tumors in \figref{fig:attributes}(a). Obviously, all tumors are randomly distributed on the lung regions without bias, indicating the universal property of PCLT20K. Besides, we show the distribution of tumor sizes using height and width in \figref{fig:attributes}(b). 
This representation highlights that the tumors within the PCLT20K are generally small, reaffirming the dataset’s value in representing typical clinical scenarios where tumors may be of minimal size.

\section{Proposed Method}
\subsection{Preliminaries}
The recent State Space Models (SSMs) \cite{gu2021combining,gu2021efficiently,smith2022simplified} represent a class of sequence-to-sequence modeling systems characterized by constant dynamics over time, a property also known as linear time-invariant (LTI). It maps a 1D sequential input $x(t) \in \mathbb{R} \rightarrow y(t) \in \mathbb{R}$ through an implicit latent state $h(t) \in \mathbb{R}^N$, which effectively capture the inherent dynamics of systems. An SSM is defined by four parameters $(\Delta, A, B, C)$ with the following operations:
\begin{equation}
\begin{aligned}
h_t  =\bar{A} h_{t-1}+\bar{B} x_t, 
y_t  =C h_t,
\end{aligned}
\end{equation}
where $(\bar{A}, \bar{B})$ represent the zero-order hold (ZOH) \cite{gu2021efficiently} discretization of $(A, B)$ using the transformations $\bar{A}=\exp (\Delta A)$ and $ \bar{B}=(\Delta A)^{-1}(\exp (A)-I)\cdot \Delta B$. This discretization facilitates efficient parallelizable training through a global convolution approach:
\begin{equation}
\begin{aligned}
 y    = x   * \overline{ K  }, 
\overline{ K  }  =\left( C   \overline{ B}  ,  C   \overline{ A B  }, \ldots,  C   \overline{ A  }^{L-1} \overline{ B  }\right),
\end{aligned}
\end{equation}
where $*$ represents the convolution operation, $\overline{K} \in \mathbb{R}^L$ is the SSM kernel.
Although discretization enhances computational efficiency, the parameters $(\Delta, \bar{A}, \bar{B}, C)$ in the SSM are data-independent and time-invariant, limiting the expressiveness of the hidden state to compress seen information. Selective SSM (or Mamba) \cite{gu2023mamba} introduces data-dependent parameters $(B, C, \Delta)$ that effectively select relevant information in $x_t$, which vary with the input through liner projections: $ B_t=$ Linear $_B\left(x_t\right), C_t=$ $\operatorname{Linear}_C\left(x_t\right), \Delta_t=$ SoftPlus(Linear $\left.\Delta\left(x_t\right)\right)$.

Through hardware-aware optimizations, the selective SSM achieves linear computation and memory complexity with respect to sequence length, effectively compressing essential context across the entire input sequence. In visual tasks, VMamba introduced the 2D Selective Scan (SS2D) \cite{zhu2024vision}. This method preserves the integrity of 2D image structures by scanning four directed feature sequences, each independently processed within an S6 (\ie Selective Scan Space State Sequential Model) block before being integrated to create a cohesive 2D feature map.

\subsection{Network Structure}
As illustrated in \figref{fig1}, our proposed method comprises two parallel branches for extracting features from PET- and CT-modal inputs, a channel-wise feature rectification module, and a dynamic cross-modality interaction module, which form an architecture entirely composed of state space models. During the encoding phase, each branch sequentially processes the input through four Visual State Space (VSS) blocks with downsampling operations to extract multi-level image features. 
The two encoder branches share weights to optimize computational efficiency. At each stage of the dual-branch architecture, a channel-wise rectification module is introduced between both branches to refine features based on two modalities. Additionally, cross-modal feature fusion is facilitated at each stage using rectified features from distinct sources through our dynamic cross-modality interaction module. In the decoding phase, fused features are further processed using Channel-Aware Visual State Space (CVSS) blocks proposed in \cite{wan2024sigma} with upsampling operations. Finally, the resulting features are fed into a classifier to generate lesion segmentation.

\subsection{Channel-wise Rectification Module}
One limitation of the existing Mamba-based network is that, although the global context is captured by scanning image feature maps, the channel information, especially across different modalities, is usually overlooked. To address this, we propose a channel-wise rectification module (CRM) that integrates multi-modal features to learn shared representations and enhance correlated features between the two modalities.
As shown in \figref{fig1}(b), given two input features $\mathbf{X}_{PET}^{in}$ and $\mathbf{X}_{CT}^{in}$ from the PET and CT modalities, we first concatenate them along the channel dimension and apply layer normalization to obtain the feature $\mathbf{X} \in \mathbb{R}^{H \times W \times 2C}$. We transpose $\mathbf{X}$ to $\mathbf{X}^T \in \mathbb{R}^{2C \times H \times W}$ and flatten it to $\hat{\mathbf{X}}^T \in \mathbb{R}^{2C \times L}$, where $L = H \times W$. This can be interpreted as using flattened feature pixels as the channel representation. We then apply the 1D selective SSM by:
\begin{equation}
\hat{\mathbf{X}}^{T^{\prime}}=\operatorname{SelectiveSSM}\left(\hat{\mathbf{X}}^T\right).
\end{equation}

This operation effectively mixes and memorizes channel-wise correlation features by scanning the channel mapping representations of the two modalities. Next, we employ linear projection $\mathcal{F}$, followed by a sigmoid function $\sigma$ to obtain weights $\mathbf{W}^C \in \mathbb{R}^{2 C}$ for each channel within both modalities. These weights reflect the importance of each channel for the two modalities, which are split into $\mathbf{W}_{PET}^C$ and $\mathbf{W}_{CT}^C$ correspond to the PET and CT channels:
\begin{equation}
\mathbf{W}_{PET}^C, \mathbf{W}_{CT}^C=\mathcal{F}_{\text {split }}\left(\sigma\left(\mathcal{F}(\hat{\mathbf{X}}^{T^{\prime}})\right)\right. .
\end{equation}

Finally, we apply channel-wise rectification by scaling the original input features with the computed weights: 
\begin{equation}
\begin{aligned}
\mathbf{X}_{PET}^{rec}  =\mathbf{W}_{PET}^C \circledast \mathbf{X}_{PET}^{in}, 
\mathbf{X}_{CT}^{rec}  =\mathbf{W}_{CT}^C \circledast \mathbf{X}_{CT}^{in},
\end{aligned}
\end{equation}
where $\circledast$ denotes channel-wise multiplication.

By modeling dependencies among channels from both PET and CT modalities, CRM effectively integrates multi-modal information to enhance shared representation learning. This rectification helps to filter out modality-specific noise by emphasizing the most informative and correlated features between the two modalities.

\begin{figure*}[!tb]
\centering
\includegraphics[width=0.83\linewidth]{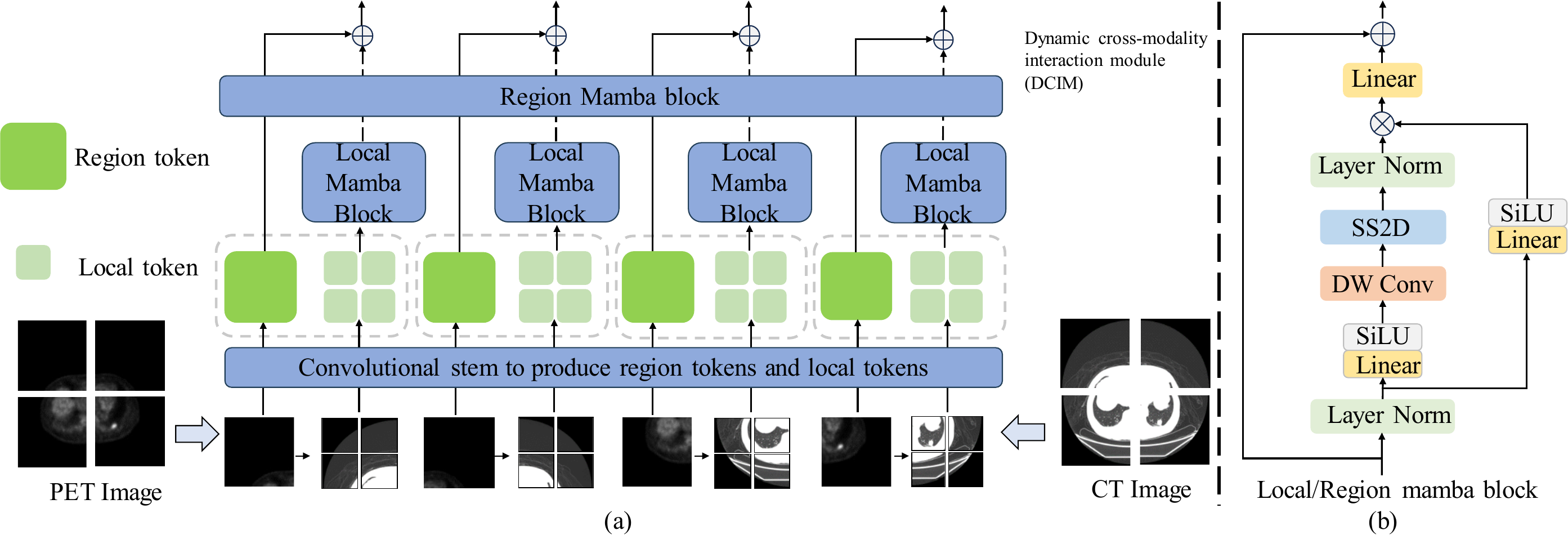}
\caption{(a) Illustration of dynamic cross-modality interaction module (DCIM), which mainly includes a convolutional stem, local Mamba block, and region Mamba block. The black dashed arrows indicate bypassing region Mamba blocks. (b) The structure of Mamba block.}
\label{fig3}
\vspace{-.1in}
\end{figure*}

\subsection{Dynamic Cross-Modality Interaction Module}
For lung PET-CT images, PET scans typically exhibit SUV values and appear brighter visually due to increased metabolic activity. However, they suffer from relatively low resolution and often present blurred tumor boundaries. Conversely, CT scans offer higher spatial resolution and provide detailed structural information about the tumor, while the contrast between the lesions and background regions is lower.
By focusing on metabolic and morphological tissue characteristics, PET and CT images can provide complementary and synergistic information for accurate tumor region identification.
However, the current VMamba \cite{zhu2024vision} cannot effectively combine this complementary regional information due to its 2D selective scan mechanism. To address this limitation, we propose a dynamic cross-modality interaction module (DCIM) to effectively integrate position and context information between PET and CT images, as shown in \figref{fig3}.

Given two rectified features $\mathbf{X}_{PET}^{rec}$ and $\mathbf{X}_{CT}^{rec}$, we first construct a convolutional stem composed of stacked $3 \times 3$ convolutions to process them and produce the PET feature $\mathbf{X}_{PET}$ $\in \mathbb{R}^{\frac{H}{r}\times \frac{W}{r} \times D} $ and CT feature $\mathbf{X}_{CT}$ $\in \mathbb{R}^{H \times W \times C}$. $r$ is the resolution of the regional patch in the PET feature, $D$ and $C$ are the feature dimensions for PET and CT, respectively. We can obtain PET regional tokens:
\begin{equation}
\mathcal{X}_{PET}=\left[X^1_{PET}, X^2_{PET}, \cdots, X^n_{PET}\right] \in \mathbb{R}^{n \times D},
\end{equation}
where $n=\frac{H}{r}\times \frac{W}{r}$ and each ${X}_{PET}^i\in \mathbb{R}^{D}$ corresponds to the PET feature in the $i$-th region.

To facilitate dynamic interactions, we first split the CT feature into $n$ regional patches:
$\mathcal{X}_{CT}=\left[X^1_{CT}, X^2_{CT}, \cdots, X^n_{CT}\right] \in \mathbb{R}^{n \times r \times r \times C}$. Each regional patch of the CT feature is further divided into $m$ local patches, i.e., a regional patch is composed of a sequence of local tokens:
\begin{equation}
X_{CT}^i = \left[x_{CT}^{i, 1}, x_{CT}^{i, 2}, \cdots, x_{CT}^{i, m}\right],
\end{equation}
where $x_{CT}^{i, j} \in \mathbb{R}^{C}$ is the $j$-th local token of the $i$-th regional patches, and $j=1,2, \cdots, m$. 

The core of our DCIM is the dynamic interaction between the position information in PET regional tokens and the structural information in CT local tokens.
Specifically, a region Mamba block is designed to process the regional position information inherent in PET features, while a local Mamba block is proposed to model the intricate dependencies among fine-grained local CT features within each regional patch. 
The region Mamba block first involves all regional tokens to learn the global position information efficiently as follows: 
\begin{equation}
\begin{aligned}
& \hat{\mathcal{X}}_{PET}=\mathcal{X}_{PET}+\operatorname{Mamba}_{region}\left(\operatorname{LN} \left(\mathcal{X}_{PET}\right)\right),
\end{aligned}
\end{equation}
where $\operatorname{LN}$ denotes layer normalization. Then the local Mamba block models the local dependencies among the local tokens within each regional patch:
\begin{equation}
\begin{aligned}
& \hat{X}_{CT}^{i}=X_{CT}^i+\operatorname{Mamba}_{local}\left( \operatorname{LN} \left(X_{CT}^i\right)\right). \\
\end{aligned}
\end{equation}

The two kinds of Mamba blocks enhance the global and local information of the tokens, respectively. To enable the high-contrast PET images to guide the high-resolution CT images, we integrate the PET regional tokens with the CT local tokens within the same region. Specifically, we add the PET regional token to each of the CT local tokens in the corresponding region:
\begin{equation}
\begin{aligned}
\footnotesize
&\hat{\mathcal{X}}_{PET}=\left[ \hat{X}^{1}_{PET}, \hat{X}^{2}_{PET}, \cdots, \hat{X}^{n}_{PET} \right] ,\\
&\hat{X}_{CT}^i =\left[ \hat{x}_{CT}^{i, 1}, \hat{x}_{CT}^{i, 2}, \cdots, \hat{x}_{CT}^{i, m} \right], \\
&{X}^i_{fuse} = \left[ \hat{x}_{CT}^{i, 1}+\hat{X}^{i}_{PET}, \hat{x}_{CT}^{i, 2}+\hat{X}^{i}_{PET}, \cdots, \hat{x}_{CT}^{i, m}+\hat{X}^{i}_{PET} \right].\\
\end{aligned}
\end{equation}

The region Mamba block focuses on extracting high-level information from regional tokens and serves as a bridge to transfer position information of lesion areas between local tokens. In this case, each local token can gather global positional information while paying close attention to its immediate neighbors.
The DCIM thus enables effective cross-modality fusion by dynamically interacting PET and CT information. It facilitates the model to leverage the complementary strengths of PET (e.g., lesion localization due to high contrast) and CT (e.g., structural details due to high resolution), allowing for a more comprehensive and synergistic feature representation that transcends traditional multiscale fusion methods.

\section{Experiments}
\label{sec:experiments}

\subsection{Implementation Details}
In this work, we implement our CIPA with the representative PyTorch platform on a PC with 4 RTX 4090 GPUs (24 GB). 
We adopt the following data enhancement: flipping horizontally or vertically, and random cropping in the size range of [0.7, 0.9]. 
We use the AdamW optimizer \cite{loshchilov2017decoupled} with an initial learning rate of $6e^{-5}$ and the cosine annealing strategy. The proposed method is trained for 50 epochs on our PCLT20K dataset with a batch size of 16. We utilize the pre-trained model on ImageNet-1K \cite{russakovsky2015imagenet} provided by VMamba \cite{zhu2024vision} for our encoder. 

We adopt four widely used metrics to evaluate the segment performance of all models, including the Intersection over Union (IoU), accuracy, F1-score, and HD95 (95\% Hausdorff Distance). 

\begin{figure*}[!tb]
\centering
\includegraphics[width=0.825\linewidth]{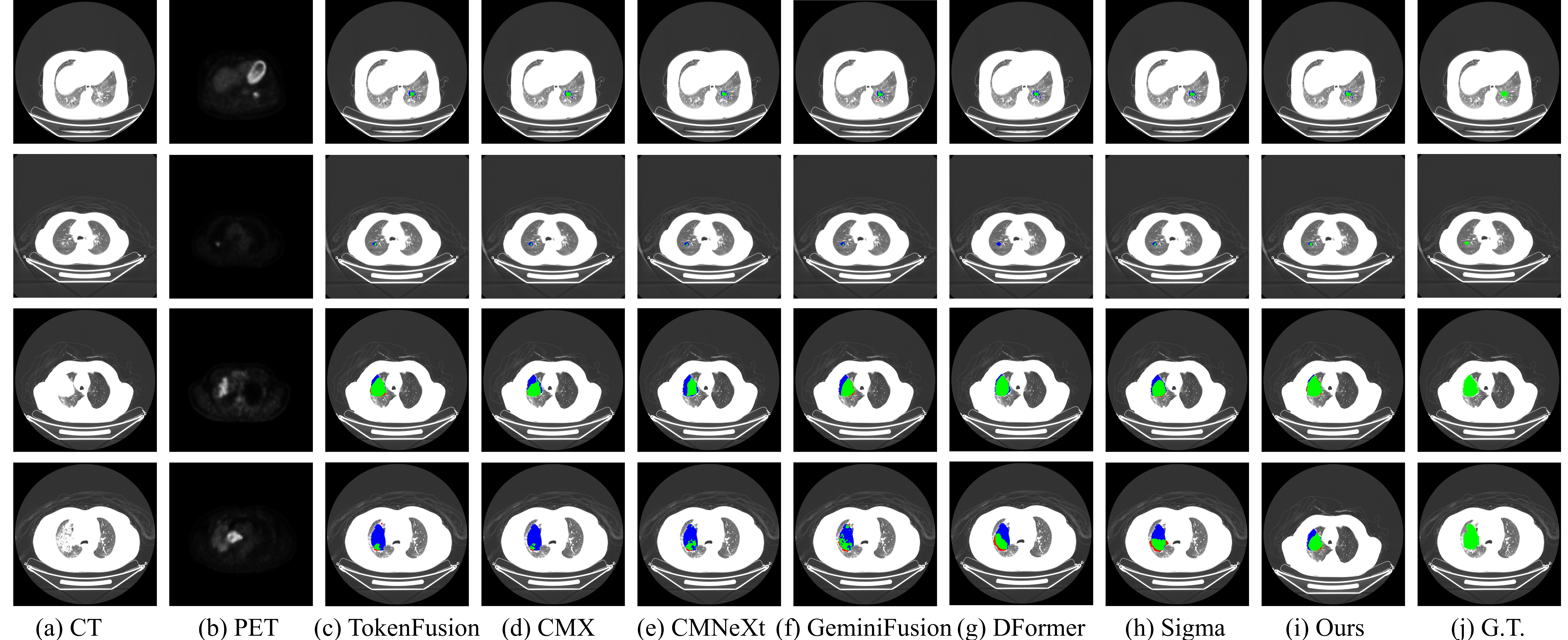}
\caption{Qualitative comparison of our CIPA with other segmentation methods on the PCLT20K dataset. Green: true positive, red: false positive, blue: false negative.}
\label{fig_comparison}
\end{figure*}

\begin{table}[!tb]
\centering
\footnotesize
\setlength\tabcolsep{1.2mm}
\begin{threeparttable}
\begin{tabular}{l|cccc|cc}
 \hline
\multirow{2}{*}{\textbf{Method}}   & \textbf{IoU $\uparrow$}  & \textbf{F1 $\uparrow$} &   \textbf{Acc $\uparrow$} & \textbf{HD95 $\downarrow$}  &  \textbf{Flops} &  \textbf{Params}
 \\
 &(\%)&(\%)&(\%)&&(G)&(M)
 \\
 \hline  
  TokenFusion \cite{wang2022multimodal}&   61.21 &75.94&   84.52& 25.52 &92.22&45.91\\
  CEN \cite{wang2022channel} &    61.28&75.99&    85.16& 30.33 &524.75&118.11\\
  CMX \cite{r65}&    61.64&76.27&    83.99& 24.68 &65.43&66.56\\
  CMNeXt \cite{zhang2023delivering}&    63.12 &77.39&   85.77& 24.95 &119.49&58.68\\
 AsymFormer \cite{du2024asymformer} &  50.03& 66.69&  76.52& 30.40 &34.20&33.05\\
  DFormer \cite{yindformer}&  61.09 &75.85&   83.83& 29.57 &14.83&39.01\\
  Sigma \cite{wan2024sigma}&   63.26&77.50&   85.37& 22.56 &76.18&48.28\\
  GeminiFusion \cite{jiageminifusion}&  62.49&76.91&   85.09& 20.70 &129.92&75.54\\
  nnU-Net \cite{isensee2021nnu} & 61.44 & 76.11 & 85.61 & 22.94& -- &--\\
  Swin-Unet \cite{cao2022swin}  & 36.41  & 53.38 & 75.48 & 124.68&62.25&54.29\\ 
  TransUNet \cite{chen2021transunet} & 51.72 & 68.18 & 78.01 & 102.97&258.58&186.46\\ 
  nnSAM \cite{li2023nnsam}  & 62.93 & 77.25 & 85.24 & 22.49&--&--\\ 
  \hline  
  CIPA (ResNet-50)  & 53.54 & 70.63 & 81.76& 38.97 & 77.77&58.04\\
  CIPA (Swin-T)  & 62.82 & 77.17 & 86.84 & 20.24 & 92.48&121.22\\
  CIPA (ours) &   \textbf{63.81} &\textbf{77.91}&    \textbf{89.01}& \textbf{17.74} &76.18&54.57\\
\hline
\end{tabular}
\end{threeparttable}
\caption{Comparison between our proposed CIPA and other methods on the testing set of our PCLT20K.}
\label{table:Comparison}
\end{table}

\subsection{Comparison with Other Methods}
\subsubsection{Quantitative Comparison}
As listed in \tabref{table:Comparison}, we compare CIPA with 12 \sArt segmentation models. The comparisons of the number of parameters and floating point operations (FLOPs) are also presented. To ensure fairness, all methods are trained on the PCLT20K-TR dataset and tested on the PCLT20K-TE dataset.
Notably, CIPA achieves the best results across all four evaluation metrics on the PCLT20K dataset. The model complexity of our CIPA is competitive compared to other methods. 
Besides, we also conduct experiments using different backbone. CIPA with Mamba as the backbone achieves superior performance with fewer parameters compared to CIPA with ResNet-50 and with Swin-T.
These superior results validate the effectiveness of CIPA for lung tumor segmentation in PET-CT images.

\subsubsection{Qualitative Comparison} 
As illustrated in \figref{fig_comparison}, we present the qualitative comparison of CIPA and segmentation methods in several examples. Overall, our method can accurately locate and segment complete lung tumors with well-preserved edge details. Additionally, CIPA produces fewer false-positive lesions, which can effectively reduce the burden on doctors during further screening. We provide additional quantitative comparison examples in the supplementary material.

\begin{table}[!tb]
\centering
\footnotesize
\setlength\tabcolsep{3.9mm}
\begin{threeparttable}
\begin{tabular}{l|ccc}
 \hline
\textbf{Method}  & \textbf{IoU (\%)}  & \textbf{F1 (\%)} &   \textbf{Acc (\%)} 
 \\
 \hline  
  TokenFusion \cite{wang2022multimodal}&   42.26&59.41&   75.52\\
  CEN \cite{wang2022channel} &    42.35&59.50&    74.97\\
  CMX \cite{r65}&    51.47&67.96&    79.66\\
  CMNeXt \cite{zhang2023delivering}&    55.75&71.59&   81.13\\
 AsymFormer \cite{du2024asymformer} &  53.07& 69.34&  80.75\\
  DFormer \cite{yindformer}&  55.99&71.78&   82.10\\
  Sigma \cite{wan2024sigma}&   58.14&73.53&   85.34\\
  GeminiFusion \cite{jiageminifusion}&  59.02 &74.23&   84.80\\
  \hline
  CIPA (ours)&   \textbf{60.33}&\textbf{75.26}&    \textbf{86.03}\\
\hline
\end{tabular}
\end{threeparttable}
\caption{Comparison between our proposed CIPA and other methods on the testing set of STS dataset \cite{vallieres2015radiomics}.}
\label{table:Comparison_sts}
\end{table}

\begin{table}[!tb]
\centering
\footnotesize
\setlength\tabcolsep{2.1 mm}
\begin{tabular}{c|cc|ccc|c}
 \hline
  \textbf{No.}&\textbf{CRM} & \textbf{DCIM} & \textbf{IoU}  & \textbf{F1} &  \textbf{Acc} &  \textbf{Params (M) }
  \\ \hline
  1& &  &62.12  & 76.64&  84.95&33.92\\
  2& \usym{1F5F8}&   &62.70 & 77.08&  86.09&36.32\\
  3& & \usym{1F5F8}&63.48 & 77.66&  86.77&52.17\\
  4& \usym{1F5F8}  & \usym{1F5F8}& \textbf{63.81}& \textbf{77.91} &  \textbf{89.01} &54.57\\ \hline
\end{tabular}
\caption{Ablation for the proposed CRM and DCIM (\%).}
\label{table:ablation_modules}
\end{table}

\begin{figure*}[ht]
\centering
\includegraphics[width=0.80\linewidth]{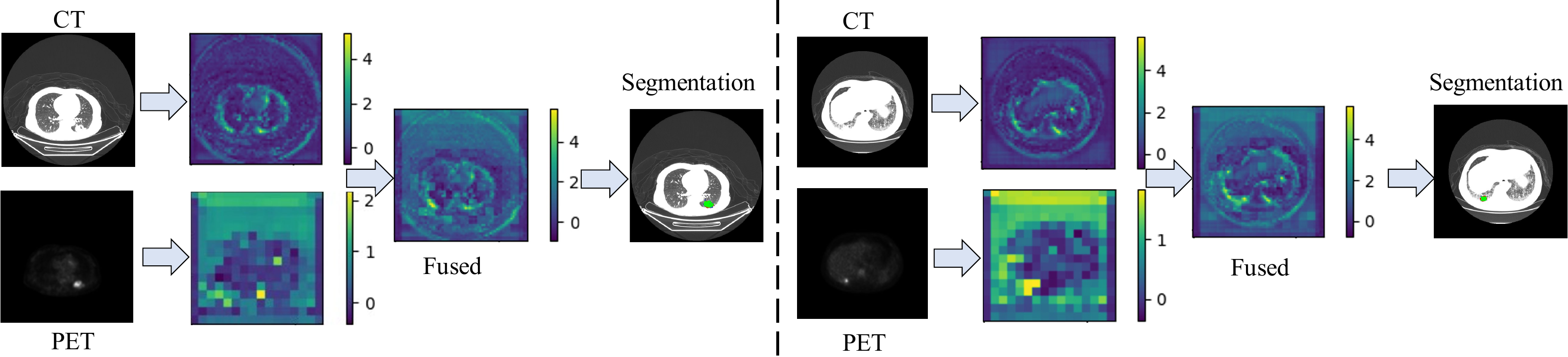}
\caption{Illustration of the process of DCIM with two examples. Each example illustrates the feature maps of the PET image processed by the region Mamba block, the feature maps of the CT image processed by the local Mamba block, and the integrated features resulting from the fusion of these two sets of information. 
}
\label{fig_DCIM}
\end{figure*}

\subsubsection{Qualitative Comparison on STS dataset} 

We also conduct comparative experiments on the publicly available PET-CT tumor segmentation dataset, STS \cite{vallieres2015radiomics}. The STS dataset consists of 2,409 pairs of PET-CT images collected from 51 patients, which are split into a training set with 1,941 pairs and a test set with 468 pairs of images. Besides, all compared methods use their default optimal experiment settings. As listed in \tabref{table:Comparison_sts}, our CIPA also achieves the best results on the STS dataset.
The \sArt results on our PCLT20K and STS datasets further validate the performance of our proposed CIPA.

\subsection{Ablation Study}
\subsubsection{Effectiveness of Key Modules}
As shown in \tabref{table:ablation_modules}, we conduct ablation experiments for the proposed CRM and DCIM in CIPA on PCLT20K dataset. By sequentially adding the CRM and DCIM modules to the baseline model based on shared Mamba, the performance across all three evaluation metrics improves. Adding both modules simultaneously achieved the best performance. These results validate the effectiveness of these two modules for tumor segmentation in PET-CT images.

\subsubsection{Effectiveness of Configurations in DCIM} 
As listed in \tabref{table:ablation_DCIM1}, we report the results for the number of local patches in the corresponding region patch in the DCIM module. It can be observed that when a PET image's regional patch contains $4 \times 4$ local patches from the CT image, both the model's complexity and segmentation performance show advantages.
Besides, we also conduct ablation studies for the various operations of DCIM in \tabref{table:ablation_DCIM2}. Compared to the five configurations of applying both region Mamba block and local Mamba block to either PET or CT image (No.1 and No.2), applying a region Mamba block to the CT image while using a local Mamba block for the PET image (No.3), applying a local Mamba block only to the CT image and directly adding the PET images (No.4), applying a region Mamba block only to the PET image and directly adding the CT images (No.5), the default configuration of DCIM (applying region Mamba block to the PET image while using local Mamba block for the CT image) achieves the best performance.

\begin{table}[!tb]
\centering
\footnotesize
\setlength\tabcolsep{2.2 mm}
\begin{tabular}{c|c|ccc|c}
 \hline
   \textbf{No.}&\textbf{ Locals in region}&\textbf{IoU}  & \textbf{F1}& \textbf{Acc} & \textbf{Params (M)}\\
   \hline
    1&2$ \times$2&62.87& 77.20& 87.59&54.92\\
   2&4$ \times$4& \textbf{63.81} & \textbf{77.91} & 89.01&54.57\\
   3&8$ \times$8& 63.07& 77.35& \textbf{89.16} &54.82\\ \hline
\end{tabular}
\caption{Ablation for the number of local patches in the corresponding region patch in DCIM (\%).}
\label{table:ablation_DCIM1}
\end{table}

\begin{table}[!tb]
\centering
\footnotesize
\setlength\tabcolsep{1.1 mm}
\begin{tabular}{c|c|ccc|c}
 \hline
   \textbf{No.}&\textbf{Operation}&\textbf{IoU}  & \textbf{F1}& \textbf{Acc}&{\textbf {Params (M) }}\\
   \hline
 1& $Region_{PET} \: \& \: Local_{PET}$& 42.06& 59.22& 80.05&54.57\\
 2&$Region_{CT} \: \& \: Local_{CT}$&61.37& 76.06& 85.28&54.57\\
 3& $Region_{CT} \: \& \: Local_{PET}$& 60.63& 75.49& 86.54&54.57\\    
   4&$Local_{CT}$&62.65& 77.04&87.68 &48.48\\
   5&$Region_{PET}$& 62.26& 76.76&86.46&48.43\\ \hline
   6&DCIM& \textbf{63.81} & \textbf{77.91} & \textbf{89.01} &54.57\\ \hline
\end{tabular}
\caption{Ablation for the various operations in DCIM (\%).}
\label{table:ablation_DCIM2}
\end{table}

\subsection{Visualization of Intermediate Feature}
To facilitate a deeper comprehension of how our proposed CIPA leverages the DCIM module to enhance the structural features within tumor areas, we visualize the features of PET image, CT image, and the features after their interaction and fusion within the DCIM, as depicted in \figref{fig_DCIM}. The results reveal that these integrated features exhibit a more pronounced delineation of lung structures and lesion areas compared to individual modalities. This improved structural clarity is advantageous for the accurate segmentation of lung tumors, thereby facilitating a more reliable basis for downstream diagnostic and therapeutic in lung oncology.

\section{Conclusion}

This paper presents a comprehensive study on lung tumor segmentation from PET-CT images. We first provide a large-scale and publicly available PET-CT lung tumor segmentation dataset, PCLT20K, which contains 21,930 pairs of PET-CT images.
Furthermore, we propose a cross-modal interactive perception network with Mamba (CIPA), which explores the innovative application of State Space Models in PET-CT tumor segmentation for the first time. A channel-wise rectification module (CRM) and a dynamic cross-modality interaction module (DCIM) are designed to effectively interact and leverage the correlation information between the multi-modal data. Extensive experiments demonstrate that CIPA achieves \sArt performance on our PCLT20K and the publicly available STS datasets. This study not only establishes a benchmark for lung tumor segmentation but also provides a solid foundation for the development of advanced multi-modal models in medical image analysis.
In the future, we plan to expand our PCLT20K dataset to support additional tasks such as PET-CT lung tumor detection and multi-modal image fusion. 

\section{Acknowledgements}
This work was supported by the National Natural Science Foundation of China under Grant 62303173, Grant 62403256, and Grant 62027810.

\newpage
{
    \small
    \bibliographystyle{ieeenat_fullname}
    \bibliography{main}

\begin{thebibliography}{45}
\providecommand{\natexlab}[1]{#1}
\providecommand{\url}[1]{\texttt{#1}}
\expandafter\ifx\csname urlstyle\endcsname\relax
  \providecommand{\doi}[1]{doi: #1}\else
  \providecommand{\doi}{doi: \begingroup \urlstyle{rm}\Url}\fi

\bibitem[Angelus and Kirby(2020)]{angelus2020large}
Pam Angelus and J Kirby.
\newblock A large-scale ct and pet/ct dataset for lung cancer diagnosis (lung-pet-ct-dx).
\newblock \emph{Cancer Imaging Archive}, 2020.

\bibitem[Cao et~al.(2022)Cao, Wang, Chen, Jiang, Zhang, Tian, and Wang]{cao2022swin}
Hu Cao, Yueyue Wang, Joy Chen, Dongsheng Jiang, Xiaopeng Zhang, Qi Tian, and Manning Wang.
\newblock Swin-unet: Unet-like pure transformer for medical image segmentation.
\newblock In \emph{Eur. Conf. Comput. Vis.}, pages 205--218. Springer, 2022.

\bibitem[Chen et~al.(2021)Chen, Lu, Yu, Luo, Adeli, Wang, Lu, Yuille, and Zhou]{chen2021transunet}
Jieneng Chen, Yongyi Lu, Qihang Yu, Xiangde Luo, Ehsan Adeli, Yan Wang, Le Lu, Alan~L Yuille, and Yuyin Zhou.
\newblock Transunet: Transformers make strong encoders for medical image segmentation.
\newblock \emph{arXiv preprint:2102.04306}, 2021.

\bibitem[Chen et~al.(2023)Chen, Qiu, Feng, and Dai]{chen2023pet}
Zhe Chen, Nan Qiu, Hui Feng, and Dongfang Dai.
\newblock Pet-ct image co-segmentation of lung tumor using joint level set model.
\newblock \emph{Comput. Electr. Eng.}, 105:\penalty0 108545, 2023.

\bibitem[Du et~al.(2024)Du, Wang, Guo, Wang, and Tang]{du2024asymformer}
Siqi Du, Weixi Wang, Renzhong Guo, Ruisheng Wang, and Shengjun Tang.
\newblock Asymformer: Asymmetrical cross-modal representation learning for mobile platform real-time rgb-d semantic segmentation.
\newblock In \emph{IEEE Conf. Comput. Vis. Pattern Recog. Worksh.}, pages 7608--7615, 2024.

\bibitem[Fu et~al.(2021)Fu, Bi, Kumar, Fulham, and Kim]{fu2021multimodal}
Xiaohang Fu, Lei Bi, Ashnil Kumar, Michael Fulham, and Jinman Kim.
\newblock Multimodal spatial attention module for targeting multimodal pet-ct lung tumor segmentation.
\newblock \emph{IEEE J. Biomed. Health Inform.}, 25\penalty0 (9):\penalty0 3507--3516, 2021.

\bibitem[Gatidis et~al.(2023)Gatidis, Fr{\"u}h, Fabritius, Gu, Nikolaou, La~Foug{\`e}re, Ye, He, Peng, Bi, et~al.]{gatidis2023autopet}
Sergios Gatidis, Marcel Fr{\"u}h, Matthias Fabritius, Sijing Gu, Konstantin Nikolaou, Christian La~Foug{\`e}re, Jin Ye, Junjun He, Yige Peng, Lei Bi, et~al.
\newblock The autopet challenge: towards fully automated lesion segmentation in oncologic pet/ct imaging.
\newblock \emph{Research Square}, 2023.

\bibitem[Gu and Dao(2023)]{gu2023mamba}
Albert Gu and Tri Dao.
\newblock Mamba: Linear-time sequence modeling with selective state spaces.
\newblock \emph{arXiv preprint arXiv:2312.00752}, 2023.

\bibitem[Gu et~al.(2021)Gu, Johnson, Goel, Saab, Dao, Rudra, and R{\'e}]{gu2021combining}
Albert Gu, Isys Johnson, Karan Goel, Khaled Saab, Tri Dao, Atri Rudra, and Christopher R{\'e}.
\newblock Combining recurrent, convolutional, and continuous-time models with linear state space layers.
\newblock \emph{Adv. Neural Inform. Process. Syst.}, 34:\penalty0 572--585, 2021.

\bibitem[Gu et~al.(2022)Gu, Goel, and Re]{gu2021efficiently}
Albert Gu, Karan Goel, and Christopher Re.
\newblock Efficiently modeling long sequences with structured state spaces.
\newblock In \emph{Int. Conf. Learn. Represent.}, 2022.

\bibitem[Huang et~al.(2024)Huang, Pei, You, Wang, Qian, and Xu]{huang2024localmamba}
Tao Huang, Xiaohuan Pei, Shan You, Fei Wang, Chen Qian, and Chang Xu.
\newblock Localmamba: Visual state space model with windowed selective scan.
\newblock \emph{arXiv preprint arXiv:2403.09338}, 2024.

\bibitem[Isensee et~al.(2021)Isensee, Jaeger, Kohl, Petersen, and Maier-Hein]{isensee2021nnu}
Fabian Isensee, Paul~F Jaeger, Simon~AA Kohl, Jens Petersen, and Klaus~H Maier-Hein.
\newblock nnu-net: a self-configuring method for deep learning-based biomedical image segmentation.
\newblock \emph{Nature methods}, 18\penalty0 (2):\penalty0 203--211, 2021.

\bibitem[Jia et~al.(2024)Jia, Guo, Han, Wu, Zhang, Xu, and Chen]{jiageminifusion}
Ding Jia, Jianyuan Guo, Kai Han, Han Wu, Chao Zhang, Chang Xu, and Xinghao Chen.
\newblock Geminifusion: Efficient pixel-wise multimodal fusion for vision transformer.
\newblock In \emph{Int. Conf. Mach. Learn.}, 2024.

\bibitem[Kumar et~al.(2019)Kumar, Fulham, Feng, and Kim]{kumar2019co}
Ashnil Kumar, Michael Fulham, Dagan Feng, and Jinman Kim.
\newblock Co-learning feature fusion maps from pet-ct images of lung cancer.
\newblock \emph{IEEE Trans. Med. Image.}, 39\penalty0 (1):\penalty0 204--217, 2019.

\bibitem[Kumar et~al.(2020)Kumar, Fulham, Feng, and Kim]{kumar2020co}
Ashnil Kumar, Michael Fulham, Dagan Feng, and Jinman Kim.
\newblock Co-learning feature fusion maps from pet-ct images of lung cancer.
\newblock \emph{IEEE Trans. Med. Image.}, 39\penalty0 (1):\penalty0 204--217, 2020.

\bibitem[Li et~al.(2020)Li, Zhao, Lu, and Tan]{li2020deep}
Laquan Li, Xiangming Zhao, Wei Lu, and Shan Tan.
\newblock Deep learning for variational multimodality tumor segmentation in pet/ct.
\newblock \emph{Neurocomputing}, 392:\penalty0 277--295, 2020.

\bibitem[Li et~al.(2023)Li, Jing, Feng, Li, He, Wang, and Zhang]{li2023nnsam}
Yunxiang Li, Bowen Jing, Xiang Feng, Zihan Li, Yongbo He, Jing Wang, and You Zhang.
\newblock nnsam: Plug-and-play segment anything model improves nnunet performance.
\newblock \emph{arXiv preprint arXiv:2309.16967}, 2023.

\bibitem[Liao et~al.(2024)Liao, Zhu, Wang, Pan, Wang, and Ma]{liao2024lightm}
Weibin Liao, Yinghao Zhu, Xinyuan Wang, Cehngwei Pan, Yasha Wang, and Liantao Ma.
\newblock Lightm-unet: Mamba assists in lightweight unet for medical image segmentation.
\newblock \emph{arXiv preprint arXiv:2403.05246}, 2024.

\bibitem[Liu et~al.(2024)Liu, Yang, Zhou, Xi, Yu, Yu, Liang, Shi, Zhang, Zheng, et~al.]{liu2024swin}
Jiarun Liu, Hao Yang, Hong-Yu Zhou, Yan Xi, Lequan Yu, Yizhou Yu, Yong Liang, Guangming Shi, Shaoting Zhang, Hairong Zheng, et~al.
\newblock Swin-umamba: Mamba-based unet with imagenet-based pretraining.
\newblock \emph{arXiv preprint arXiv:2402.03302}, 2024.

\bibitem[Loshchilov and Hutter(2017)]{loshchilov2017decoupled}
Ilya Loshchilov and Frank Hutter.
\newblock Decoupled weight decay regularization.
\newblock \emph{arXiv preprint arXiv:1711.05101}, 2017.

\bibitem[Ma and Wang(2024)]{ma2024semi}
Chao Ma and Ziyang Wang.
\newblock Semi-mamba-unet: Pixel-level contrastive and pixel-level cross-supervised visual mamba-based unet for semi-supervised medical image segmentation.
\newblock \emph{arXiv e-prints}, pages arXiv--2402, 2024.

\bibitem[Ma et~al.(2024)Ma, Li, and Wang]{ma2024u}
Jun Ma, Feifei Li, and Bo Wang.
\newblock U-mamba: Enhancing long-range dependency for biomedical image segmentation.
\newblock \emph{arXiv preprint arXiv:2401.04722}, 2024.

\bibitem[Mei et~al.(2021)Mei, Cheng, Xu, Wan, and Zhang]{mei2021sanet}
Jie Mei, Ming-Ming Cheng, Gang Xu, Lan-Ruo Wan, and Huan Zhang.
\newblock Sanet: A slice-aware network for pulmonary nodule detection.
\newblock \emph{IEEE Trans. Pattern Anal. Mach. Intell.}, 44\penalty0 (8):\penalty0 4374--4387, 2021.

\bibitem[Oreiller et~al.(2022)Oreiller, Andrearczyk, Jreige, Boughdad, Elhalawani, Castelli, Valli{\`e}res, Zhu, Xie, Peng, et~al.]{oreiller2022head}
Valentin Oreiller, Vincent Andrearczyk, Mario Jreige, Sarah Boughdad, Hesham Elhalawani, Joel Castelli, Martin Valli{\`e}res, Simeng Zhu, Juanying Xie, Ying Peng, et~al.
\newblock Head and neck tumor segmentation in pet/ct: the hecktor challenge.
\newblock \emph{Med. Image Anal.}, 77:\penalty0 102336, 2022.

\bibitem[Qiu and Xu(2022)]{qiu2022delving}
Yu Qiu and Jing Xu.
\newblock Delving into universal lesion segmentation: Method, dataset, and benchmark.
\newblock In \emph{Eur. Conf. Comput. Vis.}, pages 485--503. Springer, 2022.

\bibitem[Qiu et~al.(2022)Qiu, Liu, Li, and Xu]{qiu2022miniseg}
Yu Qiu, Yun Liu, Shijie Li, and Jing Xu.
\newblock Miniseg: an extremely minimum network based on lightweight multiscale learning for efficient covid-19 segmentation.
\newblock \emph{IEEE Trans. Neur. Net. Learn. Syst.}, 2022.

\bibitem[Russakovsky et~al.(2015)Russakovsky, Deng, Su, Krause, Satheesh, Ma, Huang, Karpathy, Khosla, Bernstein, et~al.]{russakovsky2015imagenet}
Olga Russakovsky, Jia Deng, Hao Su, Jonathan Krause, Sanjeev Satheesh, Sean Ma, Zhiheng Huang, Andrej Karpathy, Aditya Khosla, Michael Bernstein, et~al.
\newblock Imagenet large scale visual recognition challenge.
\newblock \emph{Int. J. Comput. Vis.}, 115:\penalty0 211--252, 2015.

\bibitem[Smith et~al.(2023)Smith, Warrington, and Linderman]{smith2022simplified}
Jimmy~TH Smith, Andrew Warrington, and Scott Linderman.
\newblock Simplified state space layers for sequence modeling.
\newblock In \emph{Int. Conf. Learn. Represent.}, 2023.

\bibitem[Thie(2004)]{thie2004understanding}
Joseph~A Thie.
\newblock Understanding the standardized uptake value, its methods, and implications for usage.
\newblock \emph{J. Nucl. Med.}, 45\penalty0 (9):\penalty0 1431--1434, 2004.

\bibitem[Valli{\`e}res et~al.(2015)Valli{\`e}res, Freeman, Skamene, and El~Naqa]{vallieres2015radiomics}
Martin Valli{\`e}res, Carolyn~R Freeman, Sonia~R Skamene, and Issam El~Naqa.
\newblock A radiomics model from joint fdg-pet and mri texture features for the prediction of lung metastases in soft-tissue sarcomas of the extremities.
\newblock \emph{Phys. Med. Biol.}, 60\penalty0 (14):\penalty0 5471, 2015.

\bibitem[Wan et~al.(2024)Wan, Wang, Yong, Zhang, Stepputtis, Sycara, and Xie]{wan2024sigma}
Zifu Wan, Yuhao Wang, Silong Yong, Pingping Zhang, Simon Stepputtis, Katia Sycara, and Yaqi Xie.
\newblock Sigma: Siamese mamba network for multi-modal semantic segmentation.
\newblock \emph{arXiv preprint arXiv:2404.04256}, 2024.

\bibitem[Wang et~al.(2023{\natexlab{a}})Wang, Cheng, Cao, Wu, Wang, Wei, Yan, and Liu]{wang2023mfcnet}
Fei Wang, Chao Cheng, Weiwei Cao, Zhongyi Wu, Heng Wang, Wenting Wei, Zhuangzhi Yan, and Zhaobang Liu.
\newblock Mfcnet: A multi-modal fusion and calibration networks for 3d pancreas tumor segmentation on pet-ct images.
\newblock \emph{Comput. Biol. Med.}, 155:\penalty0 106657, 2023{\natexlab{a}}.

\bibitem[Wang et~al.(2023{\natexlab{b}})Wang, Mahon, Weiss, Jan, Taylor, McDonagh, Quinn, and Yuan]{wang2023automated}
Siqiu Wang, Rebecca Mahon, Elisabeth Weiss, Nuzhat Jan, Ross~James Taylor, Philip~Reed McDonagh, Bridget Quinn, and Lulin Yuan.
\newblock Automated lung cancer segmentation using a pet and ct dual-modality deep learning neural network.
\newblock \emph{Int. J. Radiat. Oncol. Biol. Phys.}, 115\penalty0 (2):\penalty0 529--539, 2023{\natexlab{b}}.

\bibitem[Wang et~al.(2022{\natexlab{a}})Wang, Chen, Cao, Huang, Sun, and Wang]{wang2022multimodal}
Yikai Wang, Xinghao Chen, Lele Cao, Wenbing Huang, Fuchun Sun, and Yunhe Wang.
\newblock Multimodal token fusion for vision transformers.
\newblock In \emph{IEEE Conf. Comput. Vis. Pattern Recog.}, pages 12186--12195, 2022{\natexlab{a}}.

\bibitem[Wang et~al.(2022{\natexlab{b}})Wang, Sun, Huang, He, and Tao]{wang2022channel}
Yikai Wang, Fuchun Sun, Wenbing Huang, Fengxiang He, and Dacheng Tao.
\newblock Channel exchanging networks for multimodal and multitask dense image prediction.
\newblock \emph{IEEE Trans. Pattern Anal. Mach. Intell.}, 45\penalty0 (5):\penalty0 5481--5496, 2022{\natexlab{b}}.

\bibitem[Wang and Ma(2024)]{wang2024weak}
Ziyang Wang and Chao Ma.
\newblock Weak-mamba-unet: Visual mamba makes cnn and vit work better for scribble-based medical image segmentation.
\newblock \emph{arXiv preprint arXiv:2402.10887}, 2024.

\bibitem[Xiang et~al.(2022)Xiang, Zhang, Lu, and Deng]{xiang2022modality}
Dehui Xiang, Bin Zhang, Yuxuan Lu, and Shengming Deng.
\newblock Modality-specific segmentation network for lung tumor segmentation in pet-ct images.
\newblock \emph{IEEE J. Biomed. Health Inform.}, 27\penalty0 (3):\penalty0 1237--1248, 2022.

\bibitem[Xing et~al.(2024)Xing, Ye, Yang, Liu, and Zhu]{xing2024segmamba}
Zhaohu Xing, Tian Ye, Yijun Yang, Guang Liu, and Lei Zhu.
\newblock Segmamba: Long-range sequential modeling mamba for 3d medical image segmentation.
\newblock \emph{arXiv preprint arXiv:2401.13560}, 2024.

\bibitem[Yin et~al.(2024)Yin, Zhang, Li, Liu, Cheng, and Hou]{yindformer}
Bowen Yin, Xuying Zhang, Zhong-Yu Li, Li Liu, Ming-Ming Cheng, and Qibin Hou.
\newblock Dformer: Rethinking rgbd representation learning for semantic segmentation.
\newblock In \emph{Int. Conf. Learn. Represent.}, 2024.

\bibitem[Zhang et~al.(2023{\natexlab{a}})Zhang, Liu, Yang, Hu, Liu, and Stiefelhagen]{r65}
Jiaming Zhang, Huayao Liu, Kailun Yang, Xinxin Hu, Ruiping Liu, and Rainer Stiefelhagen.
\newblock Cmx: Cross-modal fusion for rgb-x semantic segmentation with transformers.
\newblock \emph{IEEE Trans. Intell. Transport. Syst.}, 2023{\natexlab{a}}.

\bibitem[Zhang et~al.(2023{\natexlab{b}})Zhang, Liu, Shi, Yang, Rei{\ss}, Peng, Fu, Wang, and Stiefelhagen]{zhang2023delivering}
Jiaming Zhang, Ruiping Liu, Hao Shi, Kailun Yang, Simon Rei{\ss}, Kunyu Peng, Haodong Fu, Kaiwei Wang, and Rainer Stiefelhagen.
\newblock Delivering arbitrary-modal semantic segmentation.
\newblock In \emph{IEEE Conf. Comput. Vis. Pattern Recog.}, pages 1136--1147, 2023{\natexlab{b}}.

\bibitem[Zhao et~al.(2024)Zhao, Huang, Tang, Li, Gao, Hu, Fan, Cheng, Yang, Zheng, et~al.]{zhao2024mmca}
Wenjie Zhao, Zhenxing Huang, Si Tang, Wenbo Li, Yunlong Gao, Yingying Hu, Wei Fan, Chuanli Cheng, Yongfeng Yang, Hairong Zheng, et~al.
\newblock Mmca-net: A multimodal cross attention transformer network for nasopharyngeal carcinoma tumor segmentation based on a total-body pet/ct system.
\newblock \emph{IEEE J. Biomed. Health Inform.}, 2024.

\bibitem[Zhong et~al.(2018)Zhong, Kim, Zhou, Plichta, Allen, Buatti, and Wu]{zhong20183d}
Zisha Zhong, Yusung Kim, Leixin Zhou, Kristin Plichta, Bryan Allen, John Buatti, and Xiaodong Wu.
\newblock 3d fully convolutional networks for co-segmentation of tumors on pet-ct images.
\newblock In \emph{IEEE ISBI}, pages 228--231. IEEE, 2018.

\bibitem[Zhong et~al.(2019)Zhong, Kim, Plichta, Allen, Zhou, Buatti, and Wu]{zhong2019simultaneous}
Zisha Zhong, Yusung Kim, Kristin Plichta, Bryan~G Allen, Leixin Zhou, John Buatti, and Xiaodong Wu.
\newblock Simultaneous cosegmentation of tumors in pet-ct images using deep fully convolutional networks.
\newblock \emph{Medical physics}, 46\penalty0 (2):\penalty0 619--633, 2019.

\bibitem[Zhu et~al.(2024)Zhu, Liao, Zhang, Wang, Liu, and Wang]{zhu2024vision}
Lianghui Zhu, Bencheng Liao, Qian Zhang, Xinlong Wang, Wenyu Liu, and Xinggang Wang.
\newblock Vision mamba: Efficient visual representation learning with bidirectional state space model.
\newblock In \emph{Int. Conf. Mach. Learn.}, 2024.

\end{thebibliography}
}

\end{document}